\documentclass[aps,twocolumn,showpacs]{revtex4}

\newcommand{\be}{\begin{equation}}
\newcommand{\ee}{  \end{equation}}
\newcommand{\ba}{\begin{eqnarray}}
\newcommand{\ea}{  \end{eqnarray}}
\newcommand{\ve}{\varepsilon}

\begin{document}

\title{Chaotic Scattering on Individual Quantum Graphs}

\author{Z. Pluha\v r$^a$ and H. A. Weidenm{\"u}ller$^b$}
\email{Hans.Weidenmueller@mpi-hd.mpg.de}
\affiliation{$^a$Faculty of Mathematics and Physics, Charles University, 180 00 Praha 8, Czech Republic \\ $^b$Max-Planck-Institut f{\"u}r Kernphysik, 69029 Heidelberg, Germany}

\begin{abstract}For chaotic scattering on quantum graphs, the
semiclassical approximation is exact. We use this fact and employ
supersymmetry, the colour-flavour transformation, and the
saddle-point approximation to calculate the exact expression for
the lowest and asymptotic expressions in the Ericson regime for
all higher correlation functions of the scattering matrix. Our
results agree with those available from the random-matrix
approach to chaotic scattering. We conjecture that our results
hold universally for quantum-chaotic scattering. 
\end{abstract}

\pacs{05.45.Mt, 03.65.Nk, 24.60.Dr}

\maketitle

\section{Introduction}

We investigate the scattering matrix ($S$ matrix) for scattering on
chaotic quantum graphs. We derive an exact expression for the lowest
$S$-matrix correlation function and asymptotic expressions for all
higher $S$-matrix correlation functions in the Ericson regime. (All
these functions are defined as wave-number averages of products of
elements of $S$.) Our motivation for this work is the following.

The celebrated conjecture by Bohigas, Giannoni and Schmit (BGS, see
Refs.~\cite{Boh84}) postulates that the spectral fluctuation
properties of closed dynamical quantum systems that are chaotic in the
classical limit, coincide with those of one of Dyson's three canonical
random-matrix ensembles~\cite{Dys62}. For the level-level correlator
of general chaotic systems with unitary or orthogonal symmetry, the
BGS conjecture has been demonstrated in Refs.~\cite{Mue04, Mue05,
  Heu07, Heu09}. For chaotic quantum graphs (which in some sense may
be considered non-Hamiltonian systems) the analogous statement was
proved in Refs.~\cite{Gnu04, Gnu05}.

For open systems, the lowest $S$-matrix correlation function plays
the same role as does the level-level correlator in closed
systems. Generalizing the BGS conjecture one may, therefore, ask
whether (depending on symmetry) that function assumes the same values
for the random-matrix approach to scattering, for scattering on
general dynamical chaotic systems, and for scattering on chaotic
quantum graphs. For the random-matrix approach, the lowest
$S$-matrix correlation function has been calculated both for the
orthogonal~\cite{Ver85} and for the unitary~\cite{Fyo05} cases. Some
higher-order correlators have also been worked out, either
exactly~\cite{Dav88, Dav89} or as leading-order terms in an
asymptotic expansion~\cite{Aga75, Wei84}. However, we are not aware of
any analytical results for the $S$-matrix correlation function either
for general dynamical chaotic systems, or for chaotic quantum
graphs. The present paper fills that gap in regard to scattering on
graphs. We show that for the case of orthogonal symmetry, the lowest
$S$-matrix correlation function for scattering on chaotic quantum
graphs coincides with the one obtained from random-matrix theory
(RMT). We also calculate asymptotic expressions for all higher
$S$-matrix correlators in the Ericson regime and show that these
coincide with RMT results if available. In the original papers on the
subject~\cite{Eri60, Eri63, Bri63} it was surmised that in the Ericson
regime, the elements of $S$ are random variables with a Gaussian
probability distribution. We show that the surmise is correct only if
the averages of all $S$-matrix elements vanish.

Chaotic scattering on quantum graphs was introduced in
Refs.~\cite{Kot00, Kot03} and extensively investigated in these papers
and in Refs.~\cite{Kot01, Sch03, Kot04} where many of its properties
were displayed with the help of numerical simulations. We benefit from
the developments in these papers and from the recent discovery of
``topological resonances'' in the chaotic scattering on
graphs~\cite{Gnu13}. We take advantage of the fact that chaotic
quantum graphs are easier to handle than general dynamical chaotic
systems because the propagator amplitudes are plane waves, and the
semiclassical approximation is exact in that case. A brief account of
this work was published in Ref.~\cite{Plu13}.

\section{Scattering on Quantum Graphs}
\label{grap}

A graph is a system of $V$ vertices labeled $\alpha, \beta, \ldots$
that are linked by $B$ bonds. We consider graphs where every vertex
$\alpha$ is linked by a single bond $(\alpha \beta)$ to every other
vertex $\beta$ (``completely connected graph''). Then the number of
bonds is $B = V (V - 1) / 2$. Our results remain valid, however, if
some bonds are missing, see Section~\ref{mass} below. The bonds
$(\alpha \beta)$ are often simply labeled with a running index $b$
that ranges from $1$ to $B$. The length of bond $b$ is denoted by
$L_b$. We assume that the lengths of all bonds are incommensurate (for
every pair $(b, b')$ of different bonds ($b \neq b'$) the ratio $L_b /
L_{b'}$ is irrational). That assumption is necessary for the graph to
be chaotic. We eventually consider the limit $V \to \infty$. We assume
that in that limit, the lengths of all bonds remain similar (i.e.,
obey $L_{\min} \leq L_b \leq L_{\rm max}$ for all $b$).

Scattering occurs when a number $\Lambda \geq 1$ of vertices is linked
by a single bond each (a ``lead'') to infinity. The lead is labeled by
the single vertex to which it is attached. The number $\Lambda$ of
leads defines the number of scattering channels and, thus, the
dimension of the scattering matrix ($S$ matrix). In the limit $V \to
\infty$ we keep the number $\Lambda$ of channels fixed so that
eventually $\Lambda \ll V$. This is analogous to the RMT approach
where the dimension $N$ of the Hamiltonian matrix is taken to infinity
while the number of channels is kept fixed. Without loss of generality
we denote the $\Lambda$ vertices coupled to a lead by $\alpha = 1,
\ldots, \Lambda$ while the indices $\alpha > \Lambda$ denote vertices
not coupled to a lead. We confine ourselves to scattering that is
time-reversal invariant.

On each bond or lead of a quantum graph, the wave function is a linear
combination of amplitudes $\exp \{ i k x \}$ and $\exp \{ - i k x \}$,
with the same wave number $k$ on all bonds or leads. The variable $x$
measures the distance to one of the vertices attached to the bond (to
the single vertex attached to the lead, respectively). The
coefficients of the linear combination are determined by boundary
conditions specified for each vertex $\alpha$ in terms of a matrix
$\Gamma^{(\alpha)}$, $\alpha = 1, \ldots, V$. The matrix
$\Gamma^{(\alpha)}$ relates outgoing amplitudes ${\cal O}$ and
incoming amplitudes ${\cal I}$ on all bonds or on the lead connected
to vertex $\alpha$ by
\be
{\cal O} = \Gamma^{(\alpha)} {\cal I} \ .
\label{1}
\ee
To ensure time-reversal invariance and flux conservation,
$\Gamma^{(\alpha)}$ is taken to be symmetric and unitary. With $\beta,
\gamma \neq \alpha$ the matrix $\Gamma^{(\alpha)}$ has the form
\ba
\Gamma^{(\alpha)} &=& \left(
  \matrix{ \rho^{(\alpha)} & \tau^{(\alpha)}_\gamma \cr
    \tau^{(\alpha)}_\beta & \sigma^{(\alpha)}_{\beta \gamma} \cr}
\right) \ {\rm for} \ \alpha \leq \Lambda \ , \nonumber \\
\Gamma^{(\alpha)} &=& \left( \sigma^{(\alpha)}_{\beta \gamma}
\right) \ {\rm for} \ \alpha > \Lambda \ .
\label{2}
\ea
Here $\rho^{(\alpha)}$ describes backscattering on lead $\alpha$,
$\tau^{(\alpha)}_\beta$ describes scattering from bond $(\alpha
\beta)$ to lead $\alpha$ or vice versa, and the $(V - 1)$-dimensional
matrix $\sigma^{(\alpha)}_{\beta \gamma}$ describes scattering from
bond $(\alpha \beta)$ to bond $(\alpha \gamma)$ or vice versa. For
$\alpha \leq \Lambda$ the symmetric matrix $\sigma^{(\alpha)}$ is
subunitary while for $\alpha > \Lambda$, it is unitary.

\section{Scattering Matrix}
\label{scat}

Given an incident wave in a single channel $\alpha \leq \Lambda$ only,
the boundary conditions~(\ref{1}) completely define the total wave
function. The amplitude of the outgoing wave in channel $\beta \leq
\Lambda$ is the element $S_{\alpha \beta}(k)$ of the symmetric and
unitary scattering matrix.

Since a full derivation of the $S$ matrix is given in
Ref.~\cite{Kot03}, we confine ourselves to a somewhat heuristic
argument. The element $S_{\alpha \beta}(k)$ of the scattering matrix
describes propagation of the wave amplitude from lead $\alpha$ to lead
$\beta$ via multiple scattering within the graph. Such multiple
scattering is governed by three elements: (i) scattering from one bond
to another by one of the matrices $\sigma^{(\alpha)}$ in
Eqs.~(\ref{2}), (ii) propagation along one of the bonds $b$ described
by the propagator $\exp \{ i k L_b \}$, (iii) scattering from a lead
to one of the bonds or vice versa by an element
$\tau^{(\alpha)}_\beta$ in Eqs.~(\ref{2}). With $\rho^{(\alpha)}$ the
amplitude for elastic backscattering on lead $\alpha$, we write
$S_{\alpha \beta}(k)$ as the sum over all ways of propagating from
lead $\alpha$ to lead $\beta$,
\ba
&& S_{\alpha \beta}(k) = \rho^{(\alpha)} \delta_{\alpha \beta} +
\tau^{(\alpha)}_\beta \exp \{ i k L_{\alpha \beta} \}
\tau^{(\beta)}_\alpha \nonumber \\
&& + \sum_\gamma \tau^{(\alpha)}_\gamma \exp \{ i k L_{\alpha \gamma}
\} \sigma^{(\gamma)}_{\alpha \beta} \exp \{ i k L_{\gamma \beta} \}
\tau^{(\beta)}_\gamma + \ldots \ . \nonumber \\
\label{2a}
\ea
The dots indicate terms of higher order in the matrices
$\sigma^{(\gamma)}$. The term containing the $n$th power of the
$\sigma^{(\gamma)}$'s is the sum of all semiclassical trajectories
that connect the vertices $\alpha$ and $\beta$ via passage through $(n
+ 1)$ bonds. Each of the traversed bonds $b$ yields the factor
$\exp \{ i k L_{b} \}$. Since $L_{\alpha \beta} = L_{\beta \alpha}$
and since $\sigma^{(\gamma)}$ is symmetric, we conclude from
Eq.~(\ref{2a}) that $S_{\alpha \beta}(k)$ is symmetric.

In order to write each term in Eq.~(\ref{2a}) as a matrix product, we
arrange the quantities mentioned under points (i) to (iii) above in
matrix form. (i) Scattering from one bond to another anywhere on the
graph is described by the matrix $\Sigma$. The non-zero elements of
$\Sigma$ are the elements of the matrices $\sigma^{(\alpha)}$. We use
two representations for $\Sigma$. In ``vertex representation'',
$\Sigma^{(V)}$ is block diagonal, each diagonal block carrying one of
the matrices $\sigma^{(\alpha)}$, $\alpha = 1, \ldots, V$. By
construction, $\Sigma^{(V)}$ has dimension $V (V - 1) = 2 B$ and is
symmetric.  Because of the factor two, it is necessary in ``bond
representation'' to double the bond label and to use directed bonds
$(b d)$. This is done by arranging the $B$ bonds $(\alpha \beta)$ in
lexicographical order (so that always $\alpha < \beta$), and by
mapping the resulting sequence onto the sequence of integers $b = 1,
\ldots, B$. To double the bond label we refer to the bonds just
constructed by a double label $(b +)$. To every such ``directed bond''
$(\alpha \beta)$ with $\alpha < \beta$ we consider the bond $(\beta
\alpha)$ and label it as $(b -)$. The number of directed bonds $(b d)$
with $d = \pm$ constructed in that way is $2 B$. In the repesentation
of directed bonds the matrix $\sigma^{(\alpha)}$ with elements
$\sigma^{(\alpha)}_{\beta \gamma}$ is mapped onto the matrix
$\sigma_{\alpha \beta, \alpha \gamma} = \sigma_{b d, b' d'}$, with the
bond labels $b$ ($b'$) determined by $(\alpha, \beta)$ (by $(\alpha,
\gamma)$, respectively), with $d = +$ ($d = -$) for $\alpha < \beta$
(for $\alpha > \beta$, respectively), and correspondingly for $d'$. In
bond representation, the matrix $\Sigma$ is written as $\Sigma^{(B)}$
and its elements are written as $\Sigma_{b d, b' d'}$. The map
$\Sigma^{(V)} \to \Sigma^{(B)}$ involves an identical rearrangement of
rows and columns. Therefore, $\Sigma^{(B)}$ is also symmetric. As for
point (iii), we arrange the elements $\tau^{(\alpha)}_\beta$ in
Eq.~(\ref{2}) that describe scattering from lead $\alpha$ to bond
$(\alpha \beta)$ in the form of a rectangular matrix ${\cal T}$ with
elements ${\cal T}_{\alpha, b d}$. The index $\alpha$ runs over all
$\Lambda$ leads, so that ${\cal T}$ has $\Lambda$ rows and $2 B$
columns. The element ${\cal T}_{\alpha, b d}$ is non-zero for every
directed bond $(b d)$ coupled to lead $\alpha$ by the element
$\tau^{(\alpha)}_\beta$, with $b$ determined by $(\alpha \beta)$ and
$d = +$ ($d = -$) for $\alpha < \beta$ (for $\alpha > \beta$,
respectively). The element $\tau^{(\beta)}_\gamma$ at the right end of
each term in Eq.~(\ref{2a}) is correspondingly written as ${\cal
T}^T_{\beta \gamma, \beta }$ with $T$ denoting the transpose. (iii)
Amplitude propagation on the directed bonds $(b d)$ is described by
the matrix $\sigma^d_1 \exp \{ i k {\cal L} \}$. In bond
representation the matrix $\exp \{ i k {\cal L} \}$ is diagonal with
elements $\delta_{b b'} \delta_{d d'} \exp \{ i k L_b \}$. For fixed
$b$ the diagonal elements are the same for $d = +$ and for $d =
-$. The matrix $\sigma^d_1$ is the direct product of the first Pauli
spin matrix in the two-dimensional space of directions $d$ and the
unit matrix in bond space. The matrix $\sigma^d_1$ reverses the
direction of all bonds. Introduction of the factor $\sigma^d_1$ is
necessary in order for the terms in Eq.~(\ref{2a}) to attain the form
of a matrix product. With these definitions, Eq.~(\ref{2a}) takes the
form
\ba
&& S_{\alpha \beta}(k) = \rho^{(\alpha)} \delta_{\alpha \beta}
\nonumber \\
&& + \sum_{b d d'} {\cal T}_{\alpha, b d} \exp \{ i k L_b \}
(\sigma^d_1)_{b d, b d'} {\cal T}^T_{b d', \beta} \nonumber \\
&& + \sum_{b d d'', b' d'} {\cal T}_{\alpha, b d} \exp \{ i k L_b \}
(\sigma^d_1)_{b d, b d''} \Sigma_{b d'', b' d'} \nonumber \\
&& \qquad \times \exp \{ i k L_{b'} \}
(\sigma^d_1)_{b' d', b' d'''} {\cal T}^T_{b' d''', \beta}
\nonumber \\
&& + \ldots \ .
\label{3}
\ea
Carrying out the summation in Eq.~(\ref{3}) we obtain
\be
S_{\alpha \beta}(k) = \rho^{(\alpha)} \delta_{\alpha \beta} + \sum_{b
d, b' d'} {\cal T}_{\alpha, b d} ({\cal W}^{-1})_{b d, b' d'}{\cal
T}^T_{b' d', \beta}
\label{4}
\ee
or in matrix notation
\be
S_{\alpha \beta}(k) = \delta_{\alpha \beta} \rho^{(\alpha)} + \big(
{\cal T} {\cal W}^{-1} {\cal T}^T \big)_{\alpha \beta}
\label{5}
\ee
where
\be
{\cal W} = \exp \{ - i k {\cal L} \} \sigma^d_1 - \Sigma^{(B)} \ .
\label{6}
\ee
Since $\exp \{ - i k {\cal L} \} \sigma^d_1$ is symmetric, so is the
matrix ${\cal W}$.  Without loss of generality we may assume that
$\rho^{(\alpha)}$ is real for all $\alpha = 1, \ldots,
\Lambda$. Indeed, for non-real $\rho^{(\alpha)}$ we write
$\rho^{(\alpha)} = \exp \{ 2 i \delta_\alpha \}
\tilde{\rho}^{(\alpha)}$ with both $\delta_\alpha$ and
$\tilde{\rho}^{(\alpha)}$ real. The transformation $S_{\alpha \beta}
\to \exp \{ - i \delta_\alpha \} S_{\alpha \beta} \exp \{ - i
\delta_\beta \}$ then removes all elastic scattering phase shifts
$\delta_\alpha$.

The $S$ matrix must be unitary and, for a time-reversal invariant
system, symmetric. Both properties follow from Eqs.~(\ref{5}) and
(\ref{6}) and from the fact that ${\cal W}$ is symmetric. We mention
in passing that our definitions differ from the ones used in
Refs.~\cite{Gnu04, Gnu05} where the factor $\sigma^d_1$ is part of the
matrix $\Sigma^{(B)}$. In Ref.~\cite{Plu13} the factor $\sigma^d_1$ in
Eq.~(\ref{6}) was erroneously omitted. That factor eventually drops
out of the calculation, however. Therefore, all results in
Ref.~\cite{Plu13} remain unchanged.

\section{Averages over the wave number $k$ and Ergodicity}

While in Hamiltonian systems averages of products of $S$-matrix
elements are taken over energy, in graph theory such averages are
taken over the wave number $k$. These are indicated by angular
brackets. In Hamiltonian systems, averages over energy of a product
containing $P \geq 1$ elements of $S$ and $Q \geq 1$ elements of $S^*$
cannot, in general, be carried out in closed form because the poles of
$S$ and those of $S^*$ lie on opposite sides of the real energy axis,
precluding the evaluation of the averages by contour integration. For
quantum graphs the exact evaluation of the average over $k$ is
possible provided the averaging interval is large compared to the
minimum difference between any two $L_b$'s. Because of the
incommensurability of the lengths $L_b$, the average over $k$ is then
equivalent~\cite{Gnu04, Gnu05} to a phase average so that for any
function $F[\exp \{ i k L_{b_1} \}, \exp \{ i k L_{b_2} \}, \ldots]$
we have
\ba
&& \langle F[\exp \{ i k L_{b_1} \}, \exp \{ i k L_{b_2} \}, \ldots]
\rangle_k \nonumber \\
&=& (1 / (2 \pi))^B \prod_{i = 1}^B \int_0^{2 \pi}
{\rm d} \phi_{b_i} F [ \exp \{ i \phi_{b_1} \}, \exp \{ i \phi_{b_2}
\}, \ldots] \nonumber \\
&=& \langle F[\exp \{ i \phi_{b_1} \}, \exp \{ i \phi_{b_2} \},
\ldots] \rangle_\phi \ .
\label{7}
\ea
The last line defines the phase average. For graphs the $B$
independent integrations over the angles $\phi_b$, $b = 1, \ldots, B$,
can be done using supersymmetry. The remarkable identity~(\ref{7})
follows~\cite{Gnu05} from an ergodicity argument. If the $L_b$ are
incommensurate, the flow in $k$ (viewed as a flow in time) on the
$B$-dimensional torus $( \exp \{ i k L_{b_1} \}, \exp \{ i k L_{b_2}
\}, \ldots)$ covers the torus densely. As the length of the averaging
interval tends to infinity, the average over ``time'' (i.e., over $k$)
can be replaced by an average over phase space (i.e., over the phases
$\phi_b$).

We compare Eq.~(\ref{7}) with the corresponding result in RMT. In RMT,
the average of an observable ${\cal O}(E)$ that depends on a
random-matrix Hamiltonian with fixed symmetry is calculated as an
ensemble average $\langle {\cal O}(E) \rangle_{\rm RM}$. The quantity
of physical interest is the average $\langle {\cal O}(E) \rangle_E$ of
${\cal O}(E)$ over energy $E$ for a given realization of the
ensemble. The equality of both averages does not hold automatically
for all observables or for all realizations of the ensemble and is
controlled by the ergodicity criterion~\cite{Bro81}
\be
\bigg\langle \bigg( \langle {\cal O}(E) \rangle_{\rm RM} - \langle {\cal
O}(E) \rangle_E \bigg)^2 \bigg\rangle_{\rm RM} = 0 \ .
\label{7a}
\ee
All terms in Eq.~(\ref{7a}) are ensemble averages or products thereof.
Hence, Eq.~(\ref{7a}) can, in principle, be tested in the framework of
RMT for every observable. If fulfilled, Eq.~(\ref{7a}) guarantees the
equality of both averages for almost all members of the ensemble,
i.e., with the exception of a set of measure zero. The excluded set
contains integrable and other Hamiltonians that do not generate RMT
fluctuations. These are so sparse that they do not contribute to the
ensemble average in Eq.~(\ref{7a}). In the case of graphs,
Eq.~(\ref{7}) unconditionally guarantees the equality of wave-number
average and phase average for all graphs with incommensurate bond
lengths. We display the close analogy between Eq.~(\ref{7}) and
Eq.~(\ref{7a}) by grouping graphs into classes. A class ${\cal C}$ is
defined by the set of all graphs with the same number $V$ of vertices
and with the same $V$ matrices $\Gamma^{(\alpha)}$ in Eqs.~(\ref{2})
that define the boundary conditions at each vertex. Graphs in ${\cal
  C}$ differ only in the lengths $L_b$ of the bonds. A given class
${\cal C}$ contains both, graphs with incommensurate and with
commensurate bond lengths.  With these definitions, the phase average
on the right-hand side of Eq.~(\ref{7}) can be read as an ensemble
average over all graphs in ${\cal C}$. Indeed, with all $L_b$ obeying
$L_{\rm min} \leq L_b \leq L_{\rm max}$, we may decrease $L_{\rm min}$
and/or increase $L_{\rm max}$ such that for fixed $k$, the quantity $k
(L_{\rm max} - L_{\rm min}) = k \Delta L$ is a multiple of $2
\pi$. Then for any function $F(\exp \{ i k L_b \})$ we have
\ba
&& \frac{1}{\Delta L} \int_{L_{\rm min}}^{L_{\rm max}} {\rm d} L_b \
F(\exp \{ i k L_b \}) \nonumber \\
&& \qquad = \frac{1}{2 \pi} \int_0^{2 \pi} {\rm d} \phi_b \ F(\exp \{ i
\phi_b \}) \ .
\label{7aa}
\ea
This statement is restricted, of course, to the $L_b$-dependence due
to the propagator amplitudes $\exp \{ i k L_b \}$. It does not apply
to any additional dependence on bond lengths $L_b$ that arises, for
instance, from the dependence of the matrices ${\cal W}$ in
Eq.~(\ref{22}) on $\kappa_p$ and $\tilde{\kappa}_q$.  Eqs.~(\ref{7})
and (\ref{7aa}) show that for every graph in ${\cal C}$ with
incommensurate bond lengths, the $k$ average $\langle ... \rangle_k$
agrees with the ensemble average $\langle ... \rangle_{ \{ L_b \} }$.
That statement can be cast into a form similar to Eq.~(\ref{7a}),
\ba
&& \bigg\langle \bigg( \big\langle F[\exp \{ i k L_{b_1} \}, \exp \{ i
k L_{b_2} \}, \ldots] \big\rangle_k \nonumber \\
&& \ - \big\langle F[\exp \{ i k L_{b_1} \}, \exp \{ i k L_{b_2} \},
\ldots] \big\rangle_{ \{ L_b \} } \bigg)^2 \bigg\rangle_{ \{ L_b \} } = 0 \ .
\nonumber \\
\label{7b}
\ea
Eq.~(\ref{7b}) holds because among the real numbers, the rational
numbers form a subset of measure zero, and the same is true of graphs
in ${\cal C}$ with commensurate bond lengths in relation to the
totality of all graphs in ${\cal C}$. Eqs.~(\ref{7a}) and (\ref{7b})
display the close similarity of the ergodicity argument for RMT and
for graphs.  The difference is that in the case of graphs, we know
analytically which graphs belong to the excluded subset of measure
zero, we know that Eq~(\ref{7}) holds strictly for graphs with
incommensurate bond lengths, and that a test of Eq.~(\ref{7b}) is,
therefore, redundant.

\section{Average $S$ Matrix}

To calculate $\langle S_{\alpha \beta} \rangle$ we use Eq.~(\ref{7})
and phase average every term of the series in Eq.~(\ref{3}). All
terms containing factors $\exp \{ i k L_b \}$ vanish and we obtain
\be
\langle S_{\alpha \beta}(k) \rangle = \delta_{\alpha \beta}
\rho^{(\alpha)} \ . 
\label{8}
\ee
As usual, we decompose $S$ into an average part and a fluctuating
part,
\be
S(k) = \langle S \rangle + S^{\rm fl}(k)
\label{9}
\ee
and have from Eqs.~(\ref{8}) and (\ref{5})
\be
S^{\rm fl}(k) = {\cal T} {\cal W}^{-1} {\cal T}^T \ .
\label{10}
\ee
It also follows from Eq.~(\ref{3}) that the average of the product of
any number of $S$-matrix elements (not containing any element $S^*$)
is equal to the product of the averages. Thus for arbitrary positive
integer $K$ and for any choice of the set $\{ \alpha_i, \beta_i, k_i
\}$ with $i = 1, \ldots, K$ we have
\be
\bigg\langle \prod_{i = 1}^K S_{\alpha_i \beta_i}(k_i) \bigg\rangle =
\prod_{i = 1}^K \bigg\langle S_{\alpha_i \beta_i} \bigg\rangle \ .
\label{11}
\ee
That same relation holds in RMT.

It is of interest to compare Eqs.~(\ref{8}) to (\ref{10}) with the
corresponding results in a Hamiltonian theory of resonance scattering.
After removal of all elastic scattering phase shifts and with $E$
denoting the energy one writes there the matrix $S$ in the
form~\cite{Mah69, Mit10}
\be
S_{\alpha \beta}(E) = \delta_{\alpha \beta} - 2 i \pi \sum_{\mu \nu}
W_{\alpha \mu} (D^{- 1}(E))_{\mu \nu} W_{\nu \beta}
\label{12}
\ee
where
\be
D_{\mu \nu}(E) = \delta_{\mu \nu} E - H_{\mu \nu} + i \pi \sum_{\gamma
= 1}^\Lambda W_{\mu \gamma} W_{\gamma \nu} \ .
\label{13}
\ee
By virtue of the coupling matrix elements $W_{\mu \alpha} = W_{\alpha
\mu}$, the eigenvalues of the $N$-dimensional Hamiltonian matrix $H$
give rise to $N$ scattering resonances. Without resonances, i.e., for
$W_{\alpha \mu} = 0$ for all $\alpha$, the $S$ matrix equals the unit
matrix. The energy average of the $S$ matrix~(\ref{12}) is~\cite{Mit10}
\be
\langle S_{\alpha \beta} \rangle_E = \delta_{\alpha \beta} \frac{1 -
x_\alpha}{1 + x_\alpha}
\label{14}
\ee 
with
\be
x_\alpha = \frac{\pi^2}{d} \frac{1}{N} \sum_{\mu = 1}^N W^2_{\alpha \mu} \ .
\label{15}
\ee
Here $d$ is the mean spacing of the resonances (of the eigenvalues of
$H$).

The standard interpretation (see Ref.~\cite{Mit10}) of $\langle S
\rangle_E$ identifies the average over energy with the fast part of
the reaction. The unitarity deficit of $\langle S \rangle_E$,
expressed in terms of the ``transmissions coefficients''
\be
T_\alpha = 1 - |\langle S_{\alpha \alpha} \rangle_E|^2 \ ,
\label{16}
\ee
measures the flux that populates the long-lived resonances of the
system. That interpretation also applies to graphs. Indeeed,
Eqs.~(\ref{8}) and (\ref{16}) and the unitarity of $\Gamma^{(\alpha)}$
give
\be
T_\alpha = \sum_\beta |\tau^{(\alpha)}_\beta |^2
\label{17}
\ee
so that $T_\alpha$ is indeed the total coupling strength connecting
lead $\alpha$ with the graph. But here the analogy between scattering
on graphs and scattering by Hamiltonian systems ends. While for fixed
average coupling strength $(1 / N) \sum_\mu W^2_{\alpha \mu}$ the
coefficient $x_\alpha$ in Eq.~(\ref{15}) changes with the density $(1
/ d)$ of the resonances (caused, for instance, by an increase of the
dimension $N$ of the Hamiltonian matrix), the analogous coefficient
$\rho^{(\alpha)}$ in Eq.~(\ref{8}) is totally independent of the
number $V$ of vertices on the graph and, thus, of the density of
resonances. The dependence of $x_\alpha$ on $1 / d$ is intuitively
understood as due to the fact that the sum of the eigenphases of $S$
increases by $\pi$ over the width of a resonance.  Averaging over the
ensuing motion of the elements of $S$ in the complex plane yields a
subunitary average $S$ matrix. The complete lack of any dependence of
$\langle S_{\alpha \alpha} \rangle$ in Eq.~(\ref{8}) on $1 / d$ is
formally due to the fact that whenever there is scattering from vertex
$\alpha$ onto any bond in the graph, propagation along that bond
causes the resulting contribution to $\langle S \rangle$ to vanish. In
physical terms that lack implies that every resonance on the graph
gives on average a vanishing contribution. In view of these distinct
differences it is not obvious that the fluctuation properties of the
$S$ matrix for chaotic scattering on graphs and for a random-matrix
model of the Hamiltonian $H$ in Eq.~(\ref{13}) should coincide.

\section{Generating Function}

With all moments of $S$ determined by Eq.~(\ref{11}) we turn to the
calculation of moments involving elements of both $S$ and $S^*$. In
view of the decomposition~(\ref{9}) all such moments can be expressed
in terms of the $(P, Q)$ correlation functions of the fluctuating part
of the $S$ matrix, defined as the average of a product of $P$ elements
of $S^{\rm fl}$ with arguments $k + \kappa_p$, $p = 1, \ldots, P$ and
$Q$ elements of $S^{{\rm fl} *}$ with arguments $k -
\tilde{\kappa}_q$, $q = 1, \ldots, Q$. Without loss of generality we
assume $P \geq Q \geq 1$.  In view of Eq.~(\ref{10}) it suffices to
work out the $(P, Q)$ correlation function of ${\cal W}^{- 1}$ defined
as
\be
\bigg\langle \prod_{p = 1}^P {\cal W}^{-1}_{b_p d_p, b'_p d'_p}(k +
\kappa_p) \prod_{q = 1}^Q \big({\cal W}^{-1}_{b_q d_q, b'_q d'_q}(k
- \tilde{\kappa}_q)\big)^* \bigg\rangle \ .
\label{22}
\ee
Let $A$ denote a symmetric matrix in directed bond space that has
non-zero elements only in the positions $(b_p d_p, b'_p d'_p)$,
\be
A_{b d, b' d'} = \delta_{b b_p} \delta_{b' b'_p} \delta_{d d_p}
\delta_{d' d'_p} + \delta_{b b'_p} \delta_{b' b_p} \delta_{d d'_p}
\delta_{d' d_p}\ .
\label{21}
\ee
We use the identity
\be
{\cal W}^{- 1}_{b_p d_p, b'_p d'_p} = {1 \over 4} \frac{\partial}
{\partial j} \frac{\det({\cal W} + j A )}{\det({\cal W} - j A)}
\bigg|_{j = 0}
\label{23}
\ee
and write the $(P, Q)$-correlation function~(\ref{22}) as
\be
(P, Q) = \frac{1}{4^{P + Q}} \prod_{p = 1}^P \prod_{q = 1}^Q
\frac{\partial}{\partial j_p} \frac{\partial}{\partial \tilde{j}_q} G
\bigg|_{j_1 = \ldots = \tilde{j}_Q = 0} \ .
\label{24}
\ee
The generating function $G$ is defined as
\be
G = \prod_{p = 1}^P \frac{\det( {\cal W} + j_p A^{(p)} )}{\det( {\cal
W} - j_p A^{(p)} )} \prod_{q = 1}^Q \frac{\det( {\cal W}^* + \tilde{j}_q
\tilde{A}^{(q)} )}{\det( {\cal W}^* - \tilde{j}_q \tilde{A}^{(q)} )} \ ,
\label{25}
\ee
with $A^{(p)}$ and $\tilde{A}^{(q)}$ defined analogously to $A$ in
Eq.~(\ref{21}).

\section{Averaging the Generating Function}

To average $G$ we use the supersymmetry approach of Refs.~\cite{Efe83,
Ver85} in the version of Refs.~\cite{Gnu04, Gnu05}. We aim at a
representation which, although closely patterned after
Ref.~\cite{Gnu05}, is self-contained. That seems advisable because
our treatment extends that of Ref.~\cite{Gnu05} to the general $(P,
Q)$ correlation function.

For $p = 1, \ldots, P $ and $q = 1, \ldots, Q$ we define the
matrices
\be
{\cal B}^{- 1}_{p \pm} = \Sigma^{(B)} \mp j_p A^{(p)} \ , \
\tilde{{\cal B}}^{- 1}_{q \pm} = \Sigma^{(B)} \mp \tilde{j}_q
\tilde{A}^{(q)} \ .
\label{26}
\ee
In directed bond space we define
\be
T_p = \exp \{ i (k + \kappa_p) {\cal L} / 2 \} \ ,
T_q = \exp \{ i (k - \tilde{\kappa}_q) {\cal L} / 2 \} \ .
\label{27}
\ee
From Eq.~(\ref{6}) we have
\ba
{\cal W}(k + \kappa_p) \pm j_p A_p = T^*_p \sigma^d_1 ( 1 - T_p
\sigma^d_1 {\cal B}^{- 1}_{p \pm} T_p) T^*_p \ .
\label{27a}
\ea
Inserting that and the corresponding expression for ${\cal W}^*$ into
Eq.~(\ref{25}) we see that the factors $T^*_p$, $T^*_q$ and the
factors $\det \sigma^d_1$ cancel in numerator and denominator. We
obtain
\ba
G &=& \prod_{p = 1}^P \frac{\det(1 - T_p \sigma^d_1 {\cal B}^{- 1}_{p +}
T_p)}{\det(1 - T_p \sigma^d_1 {\cal B}^{- 1}_{p -} T_p )} \nonumber \\
&& \times \prod_{q = 1}^Q \frac{\det(1 - T^*_q \sigma^d_1 \tilde{\cal
B}^{- 1 *}_{q +} T^*_q)}{\det(1 - T^*_q \sigma^d_1 \tilde{\cal B}^{- 1
*}_{q -} T^*_q)} \ .
\label{25a}
\ea

To write $G$ as a superintegral, we define the $8 P B$-dimensional
supervector $\psi_+$ (the $8 B Q$-dimensional supervector $\psi_-$,
respectively), both with complex commuting ($s = 1$) or anticommuting
($s = 2$) elements. The factors $2 B P$ and $2 B Q$ account for the
dimension of directed bond space and for the occurrence of the $P$
($Q$) factors in Eq.~(\ref{25a}). A factor $2$ is due to supersymmetry
with $s = 1, 2$. Another factor $2$ is due to the additional index $x
= 1, 2$. That index is introduced as a preparatory step for the
treatment of time-reversal invariance. We combine $\psi_+$ and
$\psi_-$ into a single supervector $\psi$ and define $\tilde{\psi} =
\psi^\dag$. We write the generating function as
\be
G = \prod_{p =1}^P {\rm SDet} {\cal B}_p \prod_{q = 1}^Q {\rm SDet}
\tilde{{\cal B}}^*_q \int {\rm d} (\tilde{\psi}, \psi) \ \exp \{ -
{\cal A}(\tilde{\psi}, \psi) \}
\label{29}
\ee
where ${\rm SDet}$ denotes the superdeterminant and where
\be
{\cal A}(\tilde{\psi}, \psi) = \tilde{\psi}_+ {\cal A}_+ \psi_+ +
\tilde{\psi}_- {\cal A}_- \psi_- \ .
\label{30}
\ee
The matrices ${\cal A}_+$ (${\cal A}_-$) have dimensions $8 B P$ ($8 B
Q$, respectively) and carry the indices $\{ p b d s x \}$ ($ \{ q b d
s x \}$, respectively). Moreover, the matrices ${\cal A}_+$ and ${\cal
  A}_-$ are block diagonal and contain the $P$ matrices ${\cal A}_p$,
$p = 1, \ldots, P$ (the $Q$ matrices $\tilde{{\cal A}}_q$, $q = 1,
\ldots, Q$, respectively) as diagonal blocks. These are defined by
\be  
{\cal A}_p = \left( \matrix{ 1 & T_p \cr
                T_p & {\cal B}_p \sigma^d_1 \cr} \right) \ , \
\tilde{{\cal A}}_q = \left( \matrix{ 1 & T^*_q \cr
                T^*_q & \tilde{{\cal B}}^*_q \sigma^d_1 \cr} \right) \ . 
\label{28}
\ee
With $\sigma^s_3$ the third Pauli spin matrix in two-dimensional
superspace labeled $s = 1, 2$ and in analogy to Eqs.~(\ref{26}) the
$4B$ dimensional supermatrices ${\cal B}_p$ and $\tilde{{\cal B}}^*_q$
are defined by
\be
{\cal B}^{- 1}_p = \Sigma^{(B)} - \sigma^s_3 j_p A^{(p)} \ , \
\tilde{{\cal B}}^{- 1}_q = \Sigma^{(B)} - \sigma^s_3 \tilde{j}_q
\tilde{A}^{(q)} \ .
\label{25b}
\ee
All the matrices in Eqs.~(\ref{28}) and (\ref{25b}) are diagonal in
superspace.

To account for time-reversal invariance we define
\be
\Psi = \frac{1}{\sqrt{2}} \left( \matrix{ \psi \cr
                          \sigma^d_1 \tilde{\psi}^T \cr} \right) \ ,
\tilde{\Psi} = \frac{1}{\sqrt{2}} \bigg( \tilde{\psi} \ ,
                       \psi^T \sigma^d_1 \sigma^s_3 \bigg) \ ,
\label{31}
\ee
and, for all the supermatrices (jointly denoted by $\omega$)
introduced above,
\be
\Omega = \left( \matrix{ \omega & 0 \cr
                0 & \sigma^d_1 \omega^T \sigma^d_1 \cr} \right) \ .
\label{32}
\ee
The matrix $\sigma^d_1$ accounts for time reversal in directed bond
space $d = \pm$. The two new dimensions introduced by Eqs.~(\ref{31})
and (\ref{32}) are denoted by the index $t = 1, 2$. We note that the
dimensions of the supervectors $\psi$ and $\tilde{\psi}$ remain
unchanged. In the associated two-dimensional space we define
\be
\tau = \sigma^d_1 \left( \matrix{ 0 & \sigma^s_3 \cr
                         1 & 0 \cr} \right) \ .
\label{33}
\ee
Then, $\Psi = \tau \tilde{\Psi}^T$. The generating function is now
given by
\be
G = \prod_{p = 1}^P {\rm SDet} {\cal B}_p \prod_{q = 1}^Q {\rm SDet}
\tilde{{\cal B}}^*_q \int {\rm d} (\tilde{\psi}, \psi) \ \exp \{ -
{\cal A}(\tilde{\Psi}, \Psi) \}
\label{34}
\ee
where
\be
{\cal A}(\tilde{\Psi}, \Psi) = \tilde{\Psi}_+ {\cal A}_+ \Psi_+ +
\tilde{\Psi}_- {\cal A}_- \Psi_- \ .
\label{35}
\ee
The function ${\cal A}(\tilde{\Psi}, \Psi)$ differs from the function
${\cal A}(\tilde{\psi}, \psi)$ defined in Eq.~(\ref{30}). It also
differs from the function ${\cal A}(\tilde{\Psi}, \Psi, \tilde{Z}, Z)$
defined in Eq.~(\ref{37}) below. In the sense of Eq.~(\ref{32}), the
matrices ${\cal A}_\pm$ in Eq.~(\ref{35}) are extensions of the
corresponding matrices in Eqs.~(\ref{30}), and analogously for the
matrices in Eq.~(\ref{25b}).

We use Eq.~(\ref{7}) and write the $k$ average of $G$ as a phase
average over all $\phi_b = k L_b$. The latter average is calculated
with the help of the colour-flavour transformation of
Ref.~\cite{Zir96}. In general we have $P > Q$, and we must use that
transformation in its most general form. The integrals over products
of bond propagation amplitudes are mapped onto integrals over
supermatrices $Z$ and $\tilde{Z}$. We obtain
\ba
\langle G \rangle_\phi &=& \prod_{p = 1}^P {\rm SDet} {\cal B}_p
\prod_{q = 1}^Q {\rm SDet} \tilde{{\cal B}}^*_q \int {\rm d}
(\tilde{\psi}, \psi) \int {\rm d} (\tilde{Z}, Z) \nonumber \\
&& \times \exp \{ - {\cal A}(\tilde{\Psi}, \Psi, \tilde{Z}, Z) \}
\label{36}
\ea
where the action now has the form
\be
{\cal A}(\tilde{\Psi}, \Psi, \tilde{Z}, Z) = \tilde{\Psi}_1 A_{1 1}
\Psi_1 + \tilde{\Psi}_2 A_{2 2} \Psi_2 \ .
\label{37}
\ee
The indices are those of the auxiliary label $x = 1, 2$. The matrices
$A_{1 1}$ and $A_{2 2}$ each have dimension $8 B (P + Q)$ and, in
retarded/advanced space, are given by
\ba
A_{1 1} &=& \left( \matrix{ 1 & Z z_- \cr
                        Z^\tau z_+ & 1 \cr} \right) \ ,
\nonumber \\
A_{2 2} &=& \left( \matrix{ {\cal B}_+ \sigma^d_1 & \tilde{Z}^\tau z_- \cr
           \tilde{Z} z_+ & \tilde{\cal B}^*_- \sigma^d_1 \cr} \right) \ .
\label{38}
\ea
Here ${\cal B}_+$ ($\tilde{\cal B}^*_-$) has dimension $8 B P$ ($8 B
Q$), is block diagonal, and carries the $P$ matrices ${\cal B}_p$ (the
$Q$ matrices $\tilde{{\cal B}}^*_q$, respectively) in the diagonal
blocks. Without change of notation ${\cal B}_p$ and $\tilde{{\cal
B}}^*_q$ now denote the supermatrices obtained from the supermatrices
in Eqs.~(\ref{25b}) by the doubling of matrix dimensions in
Eq.~(\ref{32}). The matrices $z_\pm$ are block diagonal and in each
block $p = 1, \ldots, P$ or $q = 1, \ldots, Q$ given by
\be
(z_+)_ p = \exp \{ i \kappa_p {\cal L} \} \ , \ (z_-)_q = \exp \{ i
\tilde{\kappa}_q {\cal L} \} \ .
\label{39}
\ee
In the general case $P > Q$ the supermatrices $Z$ and $\tilde{Z}$ are
rectangular, $Z$ having $8 B P$ rows and $8 B Q$ columns, and
conversely for $\tilde{Z}$. Both $Z$ and $\tilde{Z}$ are diagonal in
bond space. For each bond index $b$, the bond-diagonal submatrices
$Z_b$ are normalized according to
\be
\int {\rm d} (Z_b, \tilde{Z}_b) {\rm SDet} (1 - Z_b \tilde{Z}_b) = 1
\ .
\label{40}
\ee
The integration measure is the flat Berezinian. In Boson-Fermion
block notation, $Z$ and $\tilde{Z}$ have the form
\be
Z = \left( \matrix{ Z_{B B} & Z_{B F} \cr Z_{F B} & Z_{F F} \cr}
\right) \ , \ \tilde{Z} = \left( \matrix{ \tilde{Z}_{B B} &
\tilde{Z}_{B F} \cr \tilde{Z}_{F B} & \tilde{Z}_{F F} \cr} \right) \ ,
\label{41}
\ee
with
\be
\tilde{Z}_{B B} = Z^\dag_{B B} \ {\rm and} \ \tilde{Z}_{F F} = -
Z^\dag_{F F} \ .
\label{42}
\ee
The eigenvalues of the positive definite Hermitian matrix $ Z_{BB}^{
  \dagger } Z_{BB} $ are smaller than unity.  The matrices
\be
Z^{\tau} = \tau Z^{T} \tau^{-1} \ {\rm and} \ \tilde{Z}^{\tau} =
\tau \tilde{Z}^{T} \tau^{-1}
\label{43}
\ee
are simple transforms of $Z$ and $\tilde{Z}$, respectively, with
$\tau$ defined in Eq.~(\ref{33}). In the retarded/advanced block
notation of Eq.~(\ref{38}), the matrices $Z_b$ ($\tilde{Z}_b$) occur
only in the retarded/advanced (in the advanced/retarded)
non-diagonal blocks. This directly reflects the fact that $Z_b$ and
$\tilde{Z}_b$ arise from averaging the product of two propagation
amplitudes, one from the retarded and one from the advanced diagonal
blocks, respectively.

The supervectors $\psi, \tilde{\psi}$ can be integrated out. The
prefactor $\prod_{p = 1}^P {\rm SDet} {\cal B}_p \prod_{q = 1}^Q {\rm
  SDet} \tilde{{\cal B}}^*_q$ cancels out, and the averaged generating
function is (here and in what follows, we suppress the index $\phi$ on
the averages)
\be
\langle G \rangle = \int d(\tilde Z, Z) \ e^{- {\cal A} (\tilde{Z}, Z)} \ ,
\label{44}
\ee
where ${\cal A}(\tilde Z, Z)$ denotes the action
\ba 
&& {\cal A}(\tilde Z, Z) = - {\rm STr} \ln ( 1 - Z \tilde{Z})
\nonumber \\
&& + \frac{1}{2} {\rm STr} \ln ( 1 - Z z_- Z^{\tau} z_+ )
\nonumber \\
&& + \frac{1}{2} {\rm STr} \ln ( 1 - z_+ \sigma^d_1
{\cal B}_{+}^{-1} \tilde{Z}^\tau \sigma^d_1 z_- \tilde{\cal B}_-^{-1 *}
\tilde Z )
\label{45}
\ea
and ${\rm STr}$ the supertrace. Up to this point our results are
exact.

\section{Saddle-point Approximation}

We calculate $\langle G \rangle$ using the saddle-point
approximation. We neglect small terms by putting $z_{\pm} = 1$ and
$A^{(j)} = 0, \tilde{A}^{(j)} = 0$ for all $j$. Variation of the
resulting action ${\cal A}_0$ with respect to any element of $Z$ and
the relations~(\ref{43}) yield~\cite{Gnu05} the saddle-point equation
\be
\tilde{Z} \frac{1}{1 - Z \tilde{Z}} = Z^\tau \frac{1}{1 - Z Z^\tau} \ .
\label{46}
\ee
Variation of ${\cal A}_0$ with respect to any element of $\tilde{Z}$
yields correspondingly
\ba
\frac{1}{1 - Z \tilde{Z}} Z = \frac{1}{1 - \Sigma \tilde{Z}^\tau
\Sigma \tilde{Z}} \Sigma \tilde{Z}^\tau \Sigma \ .
\label{47}
\ea
Here $\Sigma$ is block diagonal in retarded-advanced space and is a
multiple of the unit matrix in superspace. In all retarded (advanced)
blocks $\Sigma$ carries the matrix $\sigma^d_1 \Sigma^{(B)}$ (the
matrix $\sigma^d_1 (\Sigma^{(B)})^*$, respectively). As in
Eq.~(\ref{6}) the factors $\sigma^d_1$ are needed to ensure proper
matrix multiplication in directed bond space. We have used that
$(\sigma^d_1 \Sigma^{(B)})^\tau = \sigma^d_1 \Sigma^{(B)}$, see
Eq.~(\ref{43}). Eq.~(\ref{46}) implies $\tilde{Z} = Z^\tau $. The
saddle-point equations~(\ref{47}) hold if (i) $Z \Sigma = \Sigma Z$
and if (ii) $\sigma^d_1 \Sigma^{(B)} \sigma^d_1 (\Sigma^{(B)})^* = 1$.
Condition (ii) is fulfilled if and only if $\Sigma^{(V)}
(\Sigma^{(V)})^* = 1$. We recall that $\Sigma^{(V)}$ is
block-diagonal, each block carrying one of the matrices
$\sigma^{(\alpha)}$ defined in Eqs.~(\ref{2}). These obey
$\sigma^{(\alpha)} (\sigma^{(\alpha)})^* = 1$ for all $\alpha >
\Lambda$. We put $\Sigma^{(V)} (\Sigma^{(V)})^* = 1$ throughout and
account for the deviation due to the blocks $\alpha \leq \Lambda$
presently. As for condition (i), we proceed as in Refs.~\cite{Gnu04,
  Gnu05} and write the universal saddle-point solutions $Z^{\rm sp}$
and $\tilde{Z}^{\rm sp}$ as
\ba
Z^{\rm sp}_{p b d s t, q b' d' s' t'} &=& \delta_{b b'} \delta_{d
d'} Y_{p s t, q s' t'} \ , \nonumber \\
\tilde{Z}^{\rm sp}_{q b d s t, p b' d' s' t'} &=& \delta_{b b'}
\delta_{d d'} \tilde{Y}_{q s t, p s' t'} \ ,
\label{49}
\ea
where the former label $x$ is replaced by $t$. Eqs.~(\ref{49})
guarantee that condition (i) is fulfilled. In retarded-advanced block
representation, the matrix ${Y}$ ($\tilde{Y}$) has non-vanishing
elements only in the retarded-advanced block (in the
advanced-retarded block, respectively). As for $Z$, the eigenvalues
of the positive definite Hermitian matrix $ Y^\dag_{BB} Y_{BB} $ must
be smaller than unity. The matrices $Y$ and $\tilde{Y}$ are linked by
the symmetry properties~(\ref{41}) and (\ref{42}). The independent
variables in the matrices $Y$ and $\tilde{Y}$ span the saddle-point
manifold. Using $\Sigma^{(V)} (\Sigma^{(V)})^* = 1$ and
Eqs.~(\ref{49}) in Eq.~(\ref{45}) we find that the saddle-point
action ${\cal A}_0$ vanishes.

Corrections to ${\cal A}_0 = 0$ are due to deviations from
$\Sigma^{(B)} (\Sigma^{(B)})^* = 1$, and from $z_\pm = 1$. As for the
former, we consider the last term of the action~(\ref{45}) (taken at
the saddle point) for $z_\pm = 1$,
\be
\frac{1}{2} {\rm STr} \ln ( 1 - \Sigma \Sigma^* Y \tilde{Y}) \ .
\label{50}
\ee
We use the fact that in the space of directed bonds, both $Y$ and
$\tilde{Y}$ are multiples of the unit matrix. We write the
term~(\ref{50}) for every block $p$ in vertex representation. For each
diagonal block $\alpha \leq \Lambda$ in the matrix $\Sigma^{(V)}$ we
use the first of Eqs.~(\ref{2}), suppressing the index $\alpha$. With
$\rho$ real, the unitary and symmetric matrix $\Gamma$ can be
unitarily transformed into
\be
\left( \matrix{ \rho & \exp \{ i \phi_1 \} T^{1/2} & 0 \cr
\exp \{ i \phi_1 \} T^{1/2} & - \rho \exp \{ 2 i \phi_1 \} & 0 \cr
0 & 0 & \delta_{\mu \nu} \exp \{ i \phi_\mu \} \cr} \right) \ . 
\label{51}
\ee
Here $\mu, \nu = 3, \ldots, V - 1$. The transmission coefficient $T$
is defined in Eq.~(\ref{17}), the phases $\phi_1$ and $\phi_\mu$ are
real and arbitrary. Eq.~(\ref{51}) shows that $\sigma \sigma^*$
differs from the unit matrix only in the first diagonal element which
is $1 - T$. Using that fact for all $\alpha \leq \Lambda$ in the
term~(\ref{50}) and the resulting expression in the action~(\ref{45}),
we obtain in the exponent of Eq.~(\ref{44}) the ``channel-coupling
term''
\be
CC = - \frac{1}{2} \sum_{\alpha = 1}^\Lambda {\rm STr}_{p s t} \ln
\bigg( 1 + T^{(\alpha)} \frac{Y \tilde{Y}}{1 - Y \tilde{Y}} \bigg) \ .
\label{52}
\ee
The trace extends only over the indices indicated. Concerning the
deviations from $z_\pm = 1$, we expand~\cite{Gnu04} $z_\pm$ and the
action ${\cal A}$ in Eq.~(\ref{45}) around the saddle-point value
${\cal A}_0 = 0$ up to first order in $\kappa_p$ and
$\tilde{\kappa}_{q}$, putting $\Sigma \Sigma^* = 1$. With
\be
\langle d_{\rm R} \rangle = \frac{1}{\pi} \sum_b L_b
\label{53}
\ee
the average level density~\cite{Kot03}, we obtain in the exponent of
Eq.~(\ref{44}) the ``symmetry-breaking term''
\be
SB = i \pi \langle d_{\rm R} \rangle \bigg( {\rm STr}_{p s t} \kappa
\frac{1}{1 - Y \tilde{Y}} + {\rm STr}_{q s t} \tilde{\kappa}
\frac{1}{1 - \tilde{Y} Y} \bigg) \ .
\label{54}
\ee
The matrix $\kappa$ is $\delta_{s s'} \delta_{t t'} \delta_{p p'}
\kappa_p$, and correspondingly for $\tilde{\kappa}$. Collecting
results we obtain
\be
\langle G \rangle = \int {\rm d} (Y, \tilde{Y}) \bigg( \ldots \bigg)
\exp \{ CC + SB \} \ .
\label{55}
\ee
The term in big round brackets denotes the ``source terms'' (terms in
${\cal A}$ that are proportional to one of the variables $j_p,
\tilde{j}_q$ in Eq.~(\ref{26})). According to Eq.~(\ref{24}) these
terms are needed to first order in every $j_p, \tilde{j}_q$ only.
However, expanding the action~(\ref{45}) and subsequently the
exponential in Eq.~(\ref{44}) in powers of all the $j$'s creates a
multitude of terms even if we keep only terms to first order. That is
why these are not given here explicitly. For specific applications
they are worked out in Sections~\ref{two} and \ref{eri}. The
integration measure in Eq.~(\ref{55}) is again the flat Berezinian.

\section{Massive Modes}
\label{mass}

In the derivation of Eq.~(\ref{55}) we have taken into account only
the universal form~(\ref{49}) of the saddle-point solution. We have
neglected all other parts of the matrices $Z$ and $\tilde{Z}$. In
Sections~\ref{two} and \ref{eri} we show that the resulting generating
function $\langle G \rangle$ in Eq.~(\ref{55}) gives rise to universal
results for the $S$ matrix of chaotic scattering. This fact motivates
our neglect: In this paper we use graph theory as a tool to generate
universal results without resorting to random-matrix theory.

Nevertheless we must address the question whether the neglect leading
to Eq.~(\ref{55}) is justified. Technically the approximation leading
to Eq.~(\ref{55}) is referred to as the ``neglect of massive
modes''. (While the saddle-point solution corresponds to the zero
mode of the problem, other parts of the matrices $Z$ and $\tilde{Z}$
give rise to Gaussian superintegrals. The factors in the exponents
play the role of masses.) Are there special graphs for which the
universal saddle-point solution~(\ref{49}) is actually correct? Or is
the neglect of massive modes perhaps even justified for all chaotic
graphs? Ref.~\cite{Gnu13} demonstrates the existence of
non-statistical resonance scattering on chaotic quantum graphs,
refuting such hopes and underlining the need to establish the
conditions under which the massive modes can be neglected. While a
full treatment of that problem is beyond the scope of the present
paper, we offer a conjecture that is based upon the following
considerations.

For closed quantum graphs, the neglect of massive modes has been
investigated~\cite{Gnu08, Gnu10} with the help of the supersymmetry
approach. The issue was quantum ergodicity. A graph is said to be
quantum ergodic if in the semiclassical limit the moduli of the
eigenfunctions are spread uniformly over the graph. Naively one might
expect chaotic quantum graphs to be quantum ergodic. However, the
existence of scars on graphs~\cite{Sch03} refutes that expectation.
Therefore, quantum ergodicity cannot be expected to hold universally
for chaotic graphs. On the other hand, quantum ergodicity emerges in
the supersymmetry approach~\cite{Gnu08, Gnu10} as a universal property
of graphs if the massive modes are neglected. Therefore, such neglect
can apply only under special conditions.

In Refs.~\cite{Gnu08, Gnu10} such conditions have been formulated in
terms of the analogue of our matrix $\Sigma^{(B)}$. In contrast to the
present case, for closed graphs that matrix contains exclusively the
elements of the unitary matrices $\sigma^{(\alpha)}$ appearing in the
second of Eqs.~(\ref{2}). For sequences of graphs with monotonically
increasing vertex number $V$ the neglect of massive modes is
asymptotically ($V \to \infty$) justified~\cite{Gnu08, Gnu10} if the
spectrum of eigenvalues of the matrix $|(\sigma^d_1 \Sigma^{(B)})_{b
  d, b' d'}|^2$ possesses a gap separating it from zero.

In chaotic scattering the statistics of wave functions is important,
too. That fact is known from the RMT approach to chaotic
scattering~\cite{Mit10}. The eigenvalues and eigenfunctions of the
random Hamiltonian $H$ in Eqs.~(\ref{12}) and (\ref{13}) are
uncorrelated random variables. The fluctuations of the $S$ matrix are
dominated~\cite{Mit10} by the fluctuations of the eigenfunctions both
in the limit of very weakly overlapping resonances (where the
fluctuations of the eigenvalues are totally unimportant) and in the
limit of strongly overlapping resonances (where a picket-fence model
for the eigenvalues~\cite{Aga75} gives the same result as a full RMT
treatment). Conversely, deviations of the eigenfunctions from RMT
statistics may give rise to non-universal scattering. For chaotic
quantum graphs, that has been shown for ``topological
resonances''~\cite{Gnu13}. These correspond to poles of the $S$ matrix
that can be moved to the real $k$ axis by continuously changing some
bond lengths, giving rise to a bound state embedded in the
continuum. Physically, a topological resonance is related to a
``cycle'' (a closed loop on the graph). When the lengths of the bonds
forming the cycle become rationally dependent, the resulting
bound-state eigenfunction is localized on the cycle and, thus, very
far from being uniformly distributed over the graph.

In view of these facts we conjecture that a prerequisite for universal
chaotic scattering on an open chaotic quantum graph is quantum
ergodicity of the corresponding closed graph (obtained by letting all
coefficients $\tau^{(\alpha)}_\beta$ in Eqs.~(\ref{2}) tend to
zero). Quantitatively we conjecture that the criterion establishing
quantum ergodicity for closed graphs~\cite{Gnu08, Gnu10} applies
equally in the present case: The spectrum of our matrix $|(\sigma^d_1
\Sigma^{(B)})_{b d, b' d'}|^2$ (which contains $\Lambda$ subunitary
matrices $\sigma^{(\alpha)}$) should asymptotically ($V \to \infty$
with $\Lambda$ fixed) have a gap separating it from zero. The proof
would require a special investigation.

The completely connected graphs defined in Section~\ref{grap} contain
many loops and may, thus, not obey our criterion. However, with a
slight change of notation that allows for missing bonds, our
derivation also holds for other types of connected graphs.

\section{Two-point Function}
\label{two}

In the framework of the Hamiltonian approach of Eqs.~(\ref{12}) and
(\ref{13}) and with $H$ replaced by an ensemble of random matrices
with orthogonal symmetry, the $S$-matrix two-point function was
worked out in Ref.~\cite{Ver85} as an ensemble average. Using the
universal saddle-point solutions~(\ref{49}), we now calculate the
$S$-matrix two-point correlation function for a single chaotic
time-reversal invariant quantum graph and show that it coincides with
the RMT result.

We consider the average generating function $\langle G \rangle$ in
Eqs.~(\ref{55}), (\ref{52}) and (\ref{54}) for the case $(P, Q) = (1,
1)$ where the matrices $Y$ and $\tilde{Y}$ are both square matrices of
dimension four carrying indices $(s t, s' t')$. The summation over $p$
in Eq.~(\ref{52}) is redundant. In the symmetry-breaking term we put
$\kappa_1 = \kappa = \tilde{\kappa}_1$. Then
\be
SB \ (1, 1) = 2 i \pi \kappa \langle d_{\rm R} \rangle {\rm STr}_{s t}
\bigg( \frac{1}{1 - Y \tilde{Y}} \bigg) \ .
\label{56}
\ee
The part of the action~(\ref{45}) at the saddle point that is relevant
for the source terms is for $z_+ = 1 = z_-$ given by
\be
\frac{1}{2} {\rm STr} \ln ( 1 - \sigma^d_1 {\cal B}_{+}^{-1} Y
\sigma^d_1 \tilde{\cal B}_-^{-1 *} \tilde{Y}) \ .
\label{56a}
\ee
Differentiation of with respect to $j_p$ with $p = 1$ and to
$\tilde{j}_q$ with $q = 1$ yields
\ba
&& {\rm STr}_{s t} \bigg( Y \sigma^d_1 (\Sigma^{(B)})^* \tilde{Y}
\nonumber \\
&& \qquad \times \frac{1}{1 -\sigma^d_1  \Sigma^{(B)} Y \sigma^d_1
(\Sigma^{(B)})^* \tilde{Y}} \sigma^d_1 \sigma^s_3 \bigg)_{b_p d_p,
b'_p d'_p} \nonumber \\
&& \times {\rm STr}_{s t} \bigg( \tilde{Y} \frac{1}{1 - \sigma^d_1
\Sigma^{(B)} Y \sigma^d_1 (\Sigma^{(B)})^* \tilde{Y}} \nonumber \\
&& \qquad \times \sigma^d_1 \Sigma^{(B)} Y \sigma^d_1 \sigma^s_3
\bigg)_{b_q d_q, b'_q d'_q} \nonumber \\
&& + {\rm STr}_{s t} \bigg( \bigg[ \tilde{Y} \frac{1}{1 - \sigma^d_1
\Sigma^{(B)} Y \sigma^d_1 (\Sigma^{(B)})^* \tilde{Y}} \sigma^d_1
\sigma^s_3 \bigg]_{b_p d_p, b_q d_q} \nonumber \\
&& \qquad \times \bigg[ Y \frac{1}{1 - \sigma^d_1 (\Sigma^{(B)})^*
\tilde{Y} \sigma^d_1 \Sigma^{(B)} Y} \sigma^d_1 \sigma^s_3
\bigg]_{b'_p d'_p, b'_q d'_q} \bigg) \nonumber \\
&& + {\rm STr}_{s t} \bigg( \bigg[ \tilde{Y} \frac{1}{1 - \sigma^d_1
\Sigma^{(B)} Y \sigma^d_1 (\Sigma^{(B)})^* \tilde{Y}} \sigma^d_1
\sigma^s_3 \bigg]_{b_p d_p, b'_q d'_q} \nonumber \\
&& \qquad \times \bigg[ Y \frac{1}{1 - \sigma^d_1 (\Sigma^{(B)})^*
\tilde{Y} \sigma^d_1 \Sigma^{(B)} Y} \sigma^d_1 \sigma^s_3
\bigg]_{b'_p d'_p, b_q d_q} \bigg) \ . \nonumber \\
\label{57}
\ea
To calculate $\langle S^{\rm fl}_{\alpha \beta} S^{{\rm fl}*}_{\gamma
\delta} \rangle$ we use Eq.~(\ref{4}), multiply expression~(\ref{57})
with ${\cal T}_{\alpha, b_p d_p}$, with ${\cal T}^*_{\gamma, b_q
d_q}$, with ${\cal T}_{\beta, b'_p d'_p}$, and with ${\cal
T}^*_{\delta, b'_q d'_q}$, and sum over the indices $\{ b_p d_p, b'_p
d'_p, b_q d_q, b'_q d'_q\}$. We rewrite the resulting expression in
vertex space and apply the unitary transformation that brings all
matrices $\sigma^{(\alpha)}$ into the form~(\ref{51}). That same
transformation applied to the vector ${\cal T}_{\alpha, b d}$ (fixed
$\alpha$) yields a vector the components of which are all zero except
for the first one that has the value $\exp \{ i \phi^{(\alpha)}_1 \}
(T^{(\alpha)})^{1/2}$. Combining the result with Eqs.~(\ref{24}),
(\ref{52}) and (\ref{56}) we obtain
\ba
&& \langle S^{\rm fl}_{\alpha \beta} S^{{\rm fl} *}_{\gamma \delta}
\rangle = \frac{1}{16} \int {\rm d}(Y, \tilde{Y}) \ F_{\alpha \beta
\gamma \delta} \nonumber \\
&& \times \exp \bigg\{ - \frac{1}{2} \sum_{\tau = 1}^\Lambda
{\rm STr}_{s t} \ln \bigg( 1 + T^{(\tau)} \frac{Y \tilde{Y}}{1 - Y
\tilde{Y}} \bigg) \bigg\} \nonumber \\
&& \times \exp \bigg\{ 2 i \pi \kappa \langle d_{\rm R} \rangle {\rm
STr}_{s t} \bigg( \frac{1}{1 - Y \tilde{Y}} \bigg) \bigg\}
\label{58}
\ea
where
\ba
&& F_{\alpha \beta \gamma \delta} = \delta_{\alpha \beta}
\delta_{\gamma \delta} \langle S_{\alpha \alpha} \rangle T^{(\alpha)}
\langle S^*_{\gamma \gamma} \rangle T^{(\gamma)} \nonumber \\
&& \ \times {\rm STr}_{s t} \bigg( \frac{1}{1 - Y \tilde{Y} +
T^{(\alpha)} Y \tilde{Y}} Y \tilde{Y} \sigma^s_3 \bigg) \nonumber \\
&& \ \times {\rm STr}_{s t} \bigg( \frac{1}{1 - \tilde{Y} Y +
T^{(\gamma)} \tilde{Y} Y} \tilde{Y} Y  \sigma^s_3 \bigg) \nonumber \\
&& + \frac{1}{2} \bigg( \delta_{\alpha \gamma} \delta_{\beta \delta} +
\delta_{\alpha \delta} \delta_{\beta \gamma} \bigg) T^{(\alpha)}
T^{(\beta)} \nonumber \\
&& \qquad \times \bigg\{ {\rm STr}_{s t} \bigg( \frac{1}{1 - Y
\tilde{Y} + T^{(\alpha)} Y \tilde{Y}} \sigma^s_3 Y \nonumber \\
&& \qquad \times \frac{1}{1 - \tilde{Y} Y + T^{(\beta)} \tilde{Y} Y}
\sigma^s_3 \tilde{Y} \bigg) + (\alpha \leftrightarrow \beta) \bigg\} \ .
\label{59}
\ea
We note that the phases appearing in expression~(\ref{51}) have
cancelled. Eqs.~(\ref{58}) and (\ref{59}) give the two-point function
for chaotic scattering on graphs in terms of a superintegral.

It is not necessary to work out the remaining integrations because we
now show that the result in Eqs.~(\ref{58}) and (\ref{59}) coincides
with the one obtained in RMT~\cite{Ver85} where these remaining steps
have been carried out. Equations in Ref.~\cite{Ver85} are denoted by a
prefixed letter $V$. In the comparison allowance must be made for the
different definitions of the symbols ${\rm STr}$ and ${\rm trg}$ used
in the two approaches.

We recall that in retarded-advanced representation the matrices $Y$
and $\tilde{Y}$ occupy the non-diagonal blocks. We denote the
corresponding blocks of the solution of the saddle-point equation of
Ref.~\cite{Ver85} by $\sigma_{1 2}$ and $\sigma_{2 1}$. In
Ref.~\cite{Ver85} the saddle-point manifold is parametrized in the
form $T^{-1}_c \sigma_D T_c$ where the matrix $\sigma_D$ is block
diagonal and proportional to the unit matrix (minus the unit matrix)
in the retarded (the advanced block, respectively). We use
Eq.~(V.D.19) for $T_c$ and obtain 
\be
\sigma_{1 2} = t_{1 2} \sqrt{ 1 + t_{2 1} t_{1 2}} \ , \ \sigma_{2 1}
= t_{2 1} \sqrt{1 + t_{1 2} t_{2 1}} \ .
\label{60}
\ee
(A common proportionality constant has been removed by scaling). The
non-linear transformation~(\ref{60}) from the variables in $\sigma_{1
2}$ and $\sigma_{2 1}$ with a flat integration measure to the
variables in $t_{1 2}$ and $t_{2 1}$ gives rise to the non-flat
integration measure ${\rm d} \mu(t)$ appearing in Eq.~(V.7.23) and
given in Eq.~(V.8.4). We use the analogous substitutions for the
matrices $Y$ and $\tilde{Y}$,
\be
Y = t^{y}_{1 2} \sqrt{1 + t^{y}_{2 1} t^y_{1 2}} \ , \ \tilde{Y} =
t^y_{2 1} \sqrt{1 + t^y_{1 2} t^y_{2 1}} \ .
\label{61}
\ee
Since the integration measure ${\rm d} (Y, \tilde{Y})$ is flat and
since the transformations~(\ref{60}) and (\ref{61}) are identical in
form, the integration measure ${\rm d} \mu(t^y)$ for the variables in
$t^y_{1 2}$ and $t^y_{2 1}$ has the same form as ${\rm d} \mu(t)$
provided that the symmetry properties of $t^y_{1 2}$ and $t^y_{2 1}$
and of $t_{1 2}$ and $t_{2 1}$ are the same. The matrices $t^y_{1 2}$
and $t^y_{2 1}$ share the symmetry properties of $Y$ and $\tilde{Y}$
in Eqs.~(\ref{41}) and (\ref{42}). Table D.3 of Ref.~\cite{Ver85}
shows that these are also the symmetry properies of $t_{1 2}$ and
$t_{2 1}$. Therefore, a direct comparison of the two-point function
given in Eq.~(V.7.23) with that for the quantum graph (the latter
expressed in terms of $t^y_{1 2}$ and $t^y_{2 1}$) is meaningful.

With the definitions $\alpha_1 = 2 t_{1 2} t_{2 1}$ and $\alpha_2 = 2
t_{2 1} t_{1 2}$ in Eq.~(V.7.20), the channel coupling term in
Eq.~(V.7.23) coincides with the one in Eq.~(\ref{58}). In the
symmetry-breaking term in Eq.~(V.7.23) we must replace energies by
wave numbers. We identify the inverse $1 / d$ of the mean level
spacing $d$ with the average level density $\langle d_R \rangle$. The
definition~(V.3.12a) of the energy difference $\ve$ implies the
substitution $\ve \to - 2 \kappa$. Then the symmetry-breaking term in
Eq.~(V.7.23) becomes equal to the one in Eq.~(\ref{58}). For the
source terms we use the text below Eqs.~(V.3.12c) and (V.7.12)) and
identify both $I(1)$ and $I(2)$ with $\sigma^s_3 \delta_{t t'}$. Then
the source terms in Eq.~(V.7.23) become equal to the terms in
Eq.~(\ref{56}), including the numerical factors. According to
Eq.~(V.8.6b) the factor $c$ in Eq.~(V.7.23) is equal to unity.
Therefore, the entire expression~(V.7.23) coincides with our result in
Eqs.~(\ref{58}) and (\ref{59}).

We have shown that the two-point function of the $S$ matrix for a
chaotic quantum graph coincides with the RMT result. The explicit form
of that function is given in Eq.~(V.8.10) and need not be repeated
here.

\section{Ericson Regime} 
\label{eri}

Progress beyond the two-point function derived in the previous
Section is possible in the Ericson regime, defined by the condition
$\sum_\alpha T^{(\alpha)} \gg 1$. The terms of leading order in an
asymptotic expansion in inverse powers of $\sum_\alpha T^{(\alpha)}$
can be worked out for all $(P, Q)$ correlation functions.

To set the stage we first derive the asymptotic form of the two-point
function, following Ref.~\cite{Ver86}. In Eq.~(\ref{58}) we expand
both the channel-coupling term and the symmetry-breaking term in
powers of $Y \tilde{Y}$, keeping only the lowest-order terms. That
gives
\be
\exp \bigg\{ \bigg( - \frac{1}{2} \sum_{\tau = 1}^\Lambda T^{(\tau)}
+ i \pi (\kappa + \tilde{\kappa}) \langle d_R \rangle \bigg) {\rm
STr}_{s t} \bigg( Y \tilde{Y} \bigg) \bigg\} \ .
\label{62}
\ee
Because of later applications we have not put $\kappa =
\tilde{\kappa}$. The product $Y \tilde{Y}$ carries the factor
$\sum_\tau T^{(\tau)} \gg 1$. Terms of higher order in $Y \tilde{Y}$
produce higher-order terms in $(\sum_\tau T^{(\tau)})^{- 1}$ and are,
therefore, neglected. We proceed likewise in the source terms,
keeping only terms bilinear in $Y$ and $\tilde{Y}$. We obtain
\be
F_{\alpha \beta \gamma \delta} \approx \bigg( \delta_{\alpha \gamma}
\delta_{\beta \delta} + \delta_{\alpha \delta} \delta_{\beta \gamma}
\bigg) T^{(\alpha)} T^{(\beta)} {\rm STr}_{s t} \bigg( \sigma^s_3 Y
\sigma^s_3 \tilde{Y} \bigg) \ . 
\label{63}
\ee
The resulting Gaussian integrals are easily evaluated and give
\be
\langle S^{\rm fl}_{\alpha \beta}(k + \kappa) S^{{\rm fl} *}_{\gamma
\delta}(k - \tilde{\kappa}) \rangle = \frac{(\delta_{\alpha \gamma}
\delta_{\beta \delta} + \delta_{\alpha \delta} \delta_{\beta \gamma}
) T^{(\alpha)} T^{(\beta)}} {\sum_\tau T^{(\tau)} - 2 i \pi (\kappa
+ \tilde{\kappa}) \langle d_{\rm R} \rangle} \ . 
\label{64}
\ee
Replacing wave numbers by energies as in Section~\ref{two} gives
exactly the expression obtained for RMT in Refs.~\cite{Aga75, Wei84,
Ver86}.

Starting from $\langle G \rangle$ in Eq.~(\ref{55}), we use the same
approximation scheme for the general $(P, Q)$ correlation function.
The rectangular matrices $Y$ and $\tilde{Y}$ consist of blocks of
dimension four each, denoted by $Y_{p q}$ and $\tilde{Y}_{q p}$, with
elements $(Y_{p q})_{s t, s' t'} = Y_{p s t, q s' t'}$ and
$(\tilde{Y}_{q p})_{s t, s' t'} = \tilde{Y}_{q s t, p s' t'}$,
respectively. The integration measure ${\rm d}(Y, \tilde{Y})$ being
flat we have
\be
{\rm d}(Y, \tilde{Y}) = \prod_{p = 1}^P \prod_{q = 1}^Q {\rm d}(Y_{p
q}, \tilde{Y}_{q p}) \ .
\label{65}
\ee
The exponent in Eq.~(\ref{55}) is approximated by
\be
\sum_{p = 1}^P \sum_{q = 1}^Q \bigg[ - \frac{1}{2} \sum_\tau T^{(\tau)} +
i \pi \langle d_R \rangle (\kappa_p + \tilde{\kappa}_q) \bigg]
{\rm STr}_{s t} \bigg( Y_{p q} \tilde{Y}_{q p} \bigg) \ .
\label{66}
\ee
We expand the source terms (last term of the action~(\ref{45}) with
$Z$ replaced by $Y$, $\tilde{Z}$ by $\tilde{Y}$, $z_\pm$ by $1$) in
powers of $Y$ and $\tilde{Y}$, retaining only terms linear in both $Y$
and $\tilde{Y}$. Only these combine to expressions of the form
$\sum_{p q} f_p Y_{p q} f_q \tilde{Y}_{q p}$ (with some matrices $f_p$
and $f_q$) that according to Eq.~(\ref{66}) give the leading-order
contribution in the expansion in inverse powers of $\sum_\tau
T^{(\tau)}$. We obtain
\ba
&& \frac{1}{2} {\rm STr}_{p q b d s t} \ln ( 1 - \sigma^d_1 B^{- 1}_+ Y
\sigma^d_1 \tilde{B}^{- 1 *}_- \tilde{Y}) \nonumber \\
&& \approx - \frac{1}{2} \sum_{p = 1}^P \sum_{q = 1}^Q {\rm STr}_{b d s t}
\bigg( \sigma^d_1 ({\cal B}^{-1}_+)_p Y_{p q} \sigma^d_1 (\tilde{\cal
B}^{- 1 *}_-)_q \tilde{Y}_{q p} \bigg) \ . \nonumber \\ 
\label{67}
\ea
Since $P \geq Q$ we first focus on the fact that we need $P$ source
terms, each one deriving from one of the factors $({\cal B}^{-
1}_+)_p$ and carrying a different element $j_p$, $p = 1, \ldots, P$,
see Eq.~(\ref{26}). Expanding the exponential of the term~(\ref{67})
in a Taylor series we accordingly keep the term 
\be
\frac{(-)^P}{2^{P} P!} \bigg[ \sum_{p = 1}^P \sum_{q = 1}^Q {\rm STr}_{b d
s t} \bigg( - \sigma^d_1 \sigma^3_s j_p A^{(p)} Y_{p q} (\sigma^d_1
\tilde{\cal B}^{-1 *}_-)_{q} \tilde{Y}_{q p} \bigg) \bigg]^P \ .
\label{68}
\ee
Differentiation with respect to $j_p$ at $j_p = 0$ for $p = 1. \ldots,
P$ gives
\be
\frac{(-)^P}{2^P} \prod_{p = 1}^P \bigg[ \sum_{q = 1}^Q {\rm
STr}_{b d s t} \bigg( - \sigma^d_1 \sigma^s_3 A^{(p)} Y_{p q} \sigma^d_1
(\tilde{\cal B}^{-1 *}_-)_{q} \tilde{Y}_{q p} \bigg) \bigg] \ .
\label{69}
\ee
We define
\be
X_{p q} = {\rm STr}_{b d s t} \bigg( - \sigma^d_1 \sigma^s_3 A^{(p)}
Y_{p q} \sigma^d_1 (\tilde{\cal B}^{-1 *}_-)_{q} \tilde{Y}_{q p} \bigg)
\label{70}
\ee
and write expression~(\ref{69}) in the form
\be
\frac{(-)^P}{2^P} \sum_{q_1 = 1}^Q X_{1 q_1} \times \ldots \times
\sum_{q_P = 1}^Q X_{P q_P} \ .
\label{71}
\ee
Differentiation with respect to $\tilde{j}_q$ at $\tilde{j}_q = 0$ for
$q = 1, \ldots, Q$ forces $Q$ of the $P$ summation variables $q_1,
\ldots, q_P$ to take the values $1, 2, \ldots, Q$. These are denoted
by $q_{i_1}, q_{i_2}, \ldots, q_{i_Q}$. The set $\{q_{i_1}, q_{i_2},
\ldots, q_{i_Q} \}$ is a permutation of the set $\{1, 2, \ldots, Q
\}$. In the corresponding factors $X_{p q_i}$, the matrices
$(\tilde{\cal B}^{-1 *}_-)_{q_i}$ are replaced by $- \sigma^d_1
\sigma^s_3 \tilde{A}^{(q_i)}$, yielding $X_{p q_i} \to \tilde{X}_{p
  q_i} = {\rm STr} (\sigma^d_1 \sigma^s_3 A^{(p)} Y \sigma^d_1
\sigma^s_3 \tilde{A}^{(q_i)} \tilde{Y})$. For $P > Q$, $(P - Q)$
factors of the form $\sum_q X_{p q}$ remain unaffected by the
differentiation. Each of these gives $\sum_q (X_{p q})_{\tilde{j}_q =
  0}$. Changing notation we write $r_i$, $i = 1, \ldots, Q$ for those
indices $p$ that appear in one of the factors $X_{p q_i}$. The set $\{
r_1, r_2, \ldots, r_Q \}$ is a subset of $\{1, 2, \ldots, P\}$. The
remaining indices in $\{1, 2, \ldots, P\}$ are denoted by $s_i$, $i =
1, \ldots, (P - Q)$. Expression~(\ref{69}) takes the form
\be
\sum_{\rm selections} \prod_{i = 1}^{P - Q} \sum_q (X_{s_i q})_{\tilde{j}_q
= 0} \sum_{\rm permutations} \prod_{i = 1}^Q \tilde{X}_{r_i, i} \ .
\label{72}
\ee
The sum with index ``selections'' runs over all possibilities of
selecting the ${ P \choose P - Q}$ indices $s_1 < s_2 < \ldots< s_{P -
Q}$ from the set $\{1, 2, \ldots, P \}$. The sum with index
``permutations'' runs over all permutations of the remaining indices
$r_1, r_2, \ldots, r_Q$.

Combining Eqs.~(\ref{65}) and (\ref{66}) with expression~(\ref{72}) we
see that the superintegral factorizes into a product of $P Q$ terms,
each factor having the form
\be
\int {\rm d} (Y_{p q}, \tilde{Y}_{q p}) \bigg( ... \bigg) \exp \bigg\{
\bigg[ ... \bigg] {\rm STr}_{s t} \bigg( Y_{p q} \tilde{Y}_{q p} \bigg)
\bigg\} \ .
\label{73}
\ee
The content of the big straight brackets is the same as in
Eq.~(\ref{66}). The content of the big round brackets depends upon
whether the index pair $(p q)$ does or does not occur in
expression~(\ref{72}). If not, the big round brackets contain the
factor unity, and the superintegral gives unity. If it does, the big
round brackets contain the factor $\tilde{X}_{p q}$ or the factor
$(X_{p q})_{\tilde{j}_q = 0}$, as the case may be. In the first case
the Gaussian superintegral (including the normalization factor $1 /
16$ in Eq.~(\ref{24})) is the same as occurs in the derivation of
Eq.~(\ref{64}) and gives the same result, with a proper replacement of
$\kappa, \tilde{\kappa}, \alpha, \beta, \gamma, \delta$. The Gaussian
superintegral containing in the integrand the factor $(X_{p
  q})_{\tilde{j}_q = 0}$ is easily worked out. The matrix
$(\Sigma^{(B)})^*$ in $(X_{p q})_{\tilde{j}_q = 0}$ does not couple to
any channel. Therefore, the expression analogous to the
result~(\ref{64}) involves only two channels and vanishes unless both
channel indices are equal. The result is
\be
{\cal F}_{\alpha_p}(\kappa_p) = - \sum_{q = 1}^Q \frac{T^{(\alpha_p)}
\langle S_{\alpha_p \alpha_p} \rangle}{\sum_\gamma T^{(\gamma)} - 2 i
\pi (\kappa_{p} + \tilde{\kappa}_{q}) \langle d_{\rm R} \rangle}
\label{74}
\ee
where again the phase $\phi_1$ in Eq.~(\ref{51}) cancels out. The sum
over $q$ arises because in the advanced block, the matrix $\sigma^d_1
\tilde{\cal B}^{-1 *}_-$ carries the same entry $\sigma^d_1
(\Sigma^{(B)})^*$ in every subblock labeled $q$.

Collecting results we obtain
\ba
&& \bigg\langle \prod_{p = 1}^P S^{\rm fl}_{\alpha_p \beta_p}(k +
\kappa_p) \prod_{q = 1}^Q S^{{\rm fl} *}_{\alpha'_{q} \beta'_{q}}(k -
\tilde{\kappa}_{q}) \bigg\rangle \nonumber \\
&& = \sum_{\rm selections} \prod_{j = 1}^{P - Q} {\cal
F}_{\alpha_{s_j}}(\kappa_{s_j}) \nonumber \\
&& \times \sum_{\rm permutations} \prod_{q = 1}^Q \bigg\langle
S^{\rm fl}_{\overline{\alpha}_q \overline{\beta}_q}
(k + \overline{\kappa}_q) S^{{\rm fl} *}_{\alpha'_{q} \beta'_{q}}(k -
\tilde{\kappa}_{q}) \bigg\rangle \ . \nonumber \\
\label{75}
\ea
The sum labeled ``selections'' goes over all ${P \choose P - Q}$
possibilities to select $(P - Q)$ matrix elements $S^{\rm fl}$ from
the first factor on the left-hand side. These give rise to the first
product which vanishes unless every selected element is diagonal. The
remaining $Q$ elements $S^{\rm fl}$, symbolically written as $S^{\rm
  fl}_{\overline{\alpha}_q \overline{\beta}_q} (k + \overline{\kappa}_q)$,
appear as first factors in the angular brackets on the right-hand
side. The sum labeled ``${\rm permutations}$'' extends over all
permutations of these elements. Each of the terms in angular brackets
on the right-hand side is equal to the asymptotic form~(\ref{14}) of
the two-point function.

For $P \geq Q$ Eq.~(\ref{75}) gives the universal part of all $(P, Q)$
$S$ matrix correlation functions in the Ericson regime $\sum_\tau
T^{(\tau)} \gg 1$. The corresponding expressions for $P < Q$ are
obtained by complex conjugation. Combined with Eqs.~(\ref{64}) and
(\ref{74}), the universal part of the distribution of the
$k$-dependent $S$ matrix is, thus, completely known for graphs in the
Ericson regime.

We compare Eqs.~(\ref{75}), (\ref{64}) and (\ref{74}) with previous
results. All of these were derived in the framework of random-matrix
theory, most approaches using Eqs.~(\ref{12}) and (\ref{13}) and a
symmetric and real random-matrix Hamiltonian $H$. In the comparison
we use the substitutions and replacements listed in Section~\ref{two}.
We have mentioned already that the asymptotic form of the $(1, 1)$
correlation function was calculated in Refs.~\cite{Aga75, Wei84,
  Ver86}. In Ref.~\cite{Aga75} that was done with the help of a
picket-fence model for the eigenvalues of the GOE matrix $H$ in
Eq.~(\ref{13}), using the Gaussian distribution of the eigenfunctions
as the only stochastic element. The full random-matrix approach was
utilized in Ref.~\cite{Wei84}, results were obtained with the help of
the replica trick. The supersymmetry approach was used in
Ref.~\cite{Ver86}.  The results agree with each other and with our
result~(\ref{64}).  Our asymptotic result for the $(2, 2)$ correlation
function agrees with that of Ref.~\cite{Aga75} and with that of
Ref.~\cite{Wei84}. For the RMT approach the complete $(2, 1)$
correlation function was worked out in Refs.~\cite{Dav88, Dav89} with
the help of supersymmetry. Our result in Eqs.~(\ref{75}) and
(\ref{64}) agrees with the asymptotic form of the expression given
there. To the best of our knowledge, the papers cited contain the
entire available analytical information on $S$ matrix correlation
functions for systems with orthogonal symmetry. (Either of the
approaches used in Refs.~\cite{Aga75} and ~\cite{Wei84} could, in
principle, be used to calculate higher correlation functions $(P, Q)$
in the Ericson regime. In both cases the effort becomes prohibitive,
however, with increasing values of $P$ and/or $Q$.) We conclude that
within that body of information, there is complete agreement between
universal chaotic scattering on graphs and the random-matrix approach
to chaotic scattering. We expect that the same agreement would be
reached for the case of unitary symmetry, worked out in
Ref.~\cite{Fyo05} for the RMT approach, but we have not checked that.

In the early papers~\cite{Eri60, Eri63, Bri63} introducing the concept
of Ericson fluctuations, it was suggested that in the Ericson regime
the $S$-matrix elements have a Gaussian distribution. We test that
prediction against our results in Eqs.~(\ref{75}), (\ref{64}) and
(\ref{74}). For the non-diagonal elements of $S^{(\rm fl)}$ and of
$S^{({\rm fl}) *}$, Eqs.~(\ref{75}) and (\ref{74}) show that the $(P,
Q)$ correlation functions with $P \neq Q$ vanish while the $(P, P)$
correlation function has the form
\ba
&& \bigg\langle \prod_{p = 1}^P S^{\rm fl}_{\alpha_p \beta_p}(k +
\kappa_p) \prod_{q = 1}^P S^{{\rm fl} *}_{\alpha'_{q} \beta'_{q}}(k -
\tilde{\kappa}_{q}) \bigg\rangle \nonumber \\
&& = \sum_{\rm permutations} \prod_{p, q = 1}^P \bigg\langle
S^{\rm fl}_{\alpha_p \beta_p} (k + \kappa_p) S^{{\rm fl} *}_{\alpha'_{q} \beta'_{q}}(k -
\tilde{\kappa}_{q}) \bigg\rangle \ . \nonumber \\
\label{76}
\ea
The sum goes over all permutations of the elements $S^{\rm
  fl}_{\alpha_p \beta_p}$. These two features -- vanishing of all $(P,
Q)$ correlation functions with $P \neq Q$ and the form~(\ref{76}) of
the $(P, P)$ correlation function -- confirm that the non-diagonal
elements of $S^{(\rm fl)}$ and of $S^{({\rm fl}) *}$ do indeed have a
Gaussian distribution. However, the diagonal elements of $S^{(\rm
  fl)}$ and of $S^{({\rm fl}) *}$ do not share this feature. For these
elements the factors ${\cal F}_\alpha(\kappa)$ in Eq.~(\ref{74}) and
the associated factors ${\cal F}^*_\alpha(\kappa)$ do not all vanish
unless $\langle S_{\alpha \alpha} \rangle = 0$ for all $\alpha$.
Through their dependence upon the arguments $\tilde{\kappa}_q$ these
factors reflect correlations between different $S$-matrix elements.
Because of these factors, the $(P, Q)$ correlation functions do not
all vanish for $P \neq Q$, the correlation functions do not all have
the form~(\ref{76}) expected for Gaussian-distributed random
variables, and the distribution of the diagonal elements of $S^{(\rm
  fl)}$ and of $S^{({\rm fl}) *}$ is, therefore, not Gaussian. That
fact was first noted in Refs.~\cite{Dav88, Dav89}. It has a simple
explanation~\cite{Die10}. The unitarity of the scattering matrix
requires that all elements of $S$ obey $|S_{\alpha \beta}| \leq
1$. That constraint also restricts the distribution of $S$. In the
Ericson regime, the constraint is easily fulfilled for the
non-diagonal elements of $S$. All these vanish on average, and
Eq.~(\ref{64}) shows that their variances are small compared to unity
for $\sum_\tau T^{(\tau)} \gg 1$. The tails of a Gaussian centered at
zero with width as given by Eq.~(\ref{64}) are very close to zero at
the unit circle in the complex plane, and the Gaussian distribution
is, therefore, asymptotically consistent with the constraint. For
every diagonal element $S_{\alpha \alpha}$ with $\langle S_{\alpha
  \alpha} \rangle \neq 0$ the situation differs. The distribution is
now centered at $\langle S_{\alpha \alpha} \rangle$, and it is
squeezed between that center and the unit circle. That is why it
differs from a Gaussian. The squeezing becomes stronger as $|\langle
S_{\alpha \alpha} \rangle|$ increases, and one would expect that the
deviations from the Gaussian are biggest for $|\langle S_{\alpha
  \alpha} \rangle| \to 1$ whereas Eq.~(\ref{75}) shows that ${\cal
  F}^*_\alpha(\kappa)$ is biggest for $|\langle S_{\alpha \alpha}
\rangle| = 1 / \sqrt{2}$. The reason is that according to
Eq.~(\ref{9}), unitarity eventually forces $|S^{(\rm fl)}_{\alpha
  \alpha}|$ to decrease as $|\langle S_{\alpha \alpha} \rangle|$
increases so that $|S^{(\rm fl)}_{\alpha \alpha}| \to 0$ for $|\langle
S_{\alpha \alpha} \rangle| \to 1$.

The implications of deviations of the $S$-matrix distribution from
the Gaussian form for cross-section fluctuations have been discussed
in Ref.~\cite{Die10}.

\section{Summary and Conclusions} 

Following up on our earlier work~\cite{Plu13} we have in this paper
studied universal aspects of scattering on chaotic quantum graphs.
Starting point is the exact semiclassical expansion for the $S$
matrix. Using an ergodicity argument we have calculated averages of
products of $S$-matrix elements over the wave number as phase
averages. We have compared ergodicity for graphs and for RMT.

The average $S$ matrix is easy to calculate. It does not depend on the
density of resonances and, thus, differs characteristically from its
counterpart in RMT. Formal expressions for all higher $(P, Q)$
$S$-matrix correlation functions were obtained with the help of
supersymmetry, the colour-flavour transformation, and the
saddle-point approximation to the resulting superintegrals. We
conjecture that our neglect of massive modes is asymptotically
justified for sequences of graphs that when closed are quantum
ergodic. These formal results were used to calculate the $(1, 1)$
correlation function exactly. That function is equal to its
counterpart in RMT for all values of the number $\Lambda$ of
channels. We also calculated all $(P, Q)$ correlation functions in the
Ericson regime, thus determining the complete $S$-matrix distribution
function in that regime. That was done by calculating the
leading-order terms in an asymptotic expansion in inverse powers of
the sum $\sum_\tau T^{(\tau)}$ of transmission coefficients. These
terms agree with the corresponding results of RMT inasmuch as the
latter are known. In the Ericson regime the $S$ matrix has a Gaussian
distribution only if all average $S$-matrix elements vanish. The
deviations which arise otherwise are due to the unitarity constraints
on $S$.

Our results suggest several lines of future research. First, an
investigation along the lines of Refs.~\cite{Gnu08, Gnu10} is called
for to see whether our conjecture regarding the neglect of massive
modes is valid. Second, if the conjecture holds it would be
interesting to know which types of graphs obey the criterion of
asymptotic quantum ergodicity. Third, the saddle-point manifold being
known, it may be possible to obtain exact closed expressions for the
$(2, 1)$ and $(2, 2)$ correlation functions, similar in form to
Eq.~(8.10) of Ref.~\cite{Ver85}. If universally valid, these would be
useful for the analysis of intensity fluctuations of chaotic
scattering in several areas of physics.

By confining ourselves to the saddle-point solution, we have focused
attention on universal aspects of chaotic scattering on graphs. To
discuss the significance of our results for scattering on chaotic
quantum systems, we first recall the situation in closed systems.
Here, the proven equality of the level-level correlation functions
for RMT, for dynamical (i.e., Hamiltonian) chaotic quantum
systems~\cite{Mue04, Mue05, Heu07, Heu09}, and for an arbitrary
chaotic quantum graph~\cite{Gnu04, Gnu05} strongly suggests that the
spectral fluctuations in all three types of systems are identical
(including all higher correlation functions), in line with the
original BGS conjecture~\cite{Boh84}. Concerning scattering systems,
we have proved the equivalence of chaotic scattering on a single graph
and of the random-matrix model of Eqs.~(\ref{12}, \ref{13}). Our
result goes beyond that for the case of closed systems because it
encompasses not only the universal $(1, 1)$ $S$-matrix correlation
function but also the universal $(P, Q)$ correlation functions in the
Ericson regime (inasmuch as the latter are available for the
random-matrix model).  As in the case of closed systems, these facts
strongly suggest that all $S$-matrix correlation functions for both
scattering systems coincide. To complete the analogy to closed systems
it would be necessary to calculate the $(1, 1)$ $S$-matrix
correlation function for scattering on dynamical chaotic quantum
systems. It would be highly surprising if that would differ from our
and the RMT result, and similarly for higher correlation
functions. Therefore, we conjecture that scattering on graphs,
scattering on Hamiltonian chaotic quantum systems, and the RMT
approach to chaotic scattering, are completely equivalent.

ZP acknowledges support by the Czech Science Foundation under Project
No P203 - 13 - 07117S. The authors are grateful for valuable comments
to A. Altland, S. Gnutzmann, J. Kvasil, P. Cejnar, and U. Smilansky.


\end{document}